\documentclass[masmath,amssymb,aps,prb,reprint,floatfix,superscriptaddress]{revtex4-2}
\usepackage{graphicx}
\usepackage{xcolor}
\usepackage{bm}
\usepackage[version=4]{mhchem}
\newcommand{\rev}[1]{\textcolor{black}{#1}}

\begin{document}

\title{Short-range order in the CoCrFeMnNi high-entropy alloy from cluster expansion}
\author{Wei Chen}\email{wei.chen@uclouvain.be}
\affiliation{Institute of Condensed Matter and Nanoscience (IMCN), Universit\'{e} catholique de Louvain, Louvain-la-Neuve, Belgium}
\author{Gian-Marco Rignanese}
\affiliation{Institute of Condensed Matter and Nanoscience (IMCN), Universit\'{e} catholique de Louvain, Louvain-la-Neuve, Belgium}
\affiliation{WEL Research Institute, Wavre, Belgium}
\author{Geoffroy Hautier}\email{geoffroy.hautier@rice.edu}
\affiliation{Department of Materials Science and NanoEngineering, Rice University, Houston, TX, USA}
\affiliation{Rice Advanced Materials Institute, Rice University, Houston, TX, USA}
\affiliation{Thayer School of Engineering, Dartmouth College, Hanover, NH, USA}
\date{\today}

\begin{abstract}
We investigate the short-range order (SRO) and phase stability of the equiatomic CoCrFeMnNi high-entropy alloy using cluster expansion supplemented by an eigen-decomposition analysis of the SRO parameters.
Our results reveal that the primary ordering behavior is determined by strong Cr-Cr repulsive interactions, complemented by attractive heteroatomic Cr-$X$ pairs in the first nearest-neighbor shell. 
This chemical affinity is consistent with the emergence of ordered local environments and appears to be a major contributor to the primary order-disorder transition.
At lower temperatures, the spectral SRO analysis suggests two additional lower-temperature instabilities: a collective ordering instability and an Fe-rich local clustering instability.
\end{abstract}

\maketitle

\section{Introduction}

High entropy alloys (HEAs) were originally conceptualized as ideal random solid solutions stabilized by high configurational entropy~\cite{Yeh2004}. However, substantial evidence now indicates the formation of chemical short-range order (SRO) at low and intermediate temperatures, challenging the random solution assumption~\cite{Ziehl2023}. 
This local chemical ordering, which is manifested by specific preferences in atomic arrangement over only a few nearest-neighbor (NN) shells, is hypothesized to play a critical role in the extraordinary mechanical properties associated with HEAs, such as enhanced strength~\cite{Taheri2023}. 
SRO creates a rugged energy landscape that impedes dislocation motion, providing a potent strengthening mechanism distinct from traditional solid-solution hardening. 
Furthermore, SRO significantly influences key intrinsic parameters such as the stacking fault energy (SFE), which can alter deformation modes---for instance, by promoting planar slip over wavy slip---thereby impacting both strength and ductility~\cite{Cao2025}.

Accurately capturing these subtle statistical correlations requires sampling vast configurational spaces and system sizes that go beyond the computational capabilities of standard density functional theory (DFT). To overcome these limitations, we conduct a cluster expansion (CE)-based study to elucidate the order-disorder transition and the detailed nature of SRO at the atomic level in the prototypical high-entropy Cantor alloy (CoCrFeMnNi)~\cite{Cantor2004}.

\section{Methods}

In the cluster expansion formalism~\cite{Sanchez1984}, the configurational energy $E(\sigma)$ of an arbitrary atomic arrangement $\sigma$ is expressed by the linear combinations of cluster correlation functions $\Phi_\alpha(\sigma)$
\begin{equation}\label{eq:clex}
E(\sigma) = J_0 + \sum_\alpha J_\alpha \Phi_\alpha(\sigma),
\end{equation}
where the index $\alpha$ denotes a specific cluster of sites (e.g., pairs and triplets) defined on the crystal lattice, and the $J_\alpha$ are the effective cluster interactions (ECIs) to be determined. 
The cluster correlation functions $\Phi_\alpha(\sigma)$ are defined as the product of site basis functions $\Theta_{n_i}$ averaged over the entire lattice sites involved in the cluster $\alpha$
\begin{equation}\label{eq:phi}
\Phi_\alpha(\sigma) = \frac{1}{N_\text{sites} m_\alpha} \sum_{\beta \in \text{orbit}} \prod_{i \in \beta} \Theta_{n_i}(\sigma_i)
\end{equation}
where the orbit represents the set of all clusters equivalent to $\alpha$ by symmetry, $m_\alpha$ is the multiplicity of the cluster per site, and $N_\text{sites}$ is the total number of lattice sites.
For a multicomponent system with $M$ chemical components, the occupation of a lattice site $i$ is described by a set of $M-1$ orthogonal site basis functions $\Theta_{n_i}(\sigma_i)$ (Chebyshev polynomials in the present study).
In essence, cluster expansion maps the continuous configurational space onto a discrete Ising-like model, allowing for the efficient evaluation of thermodynamic properties.

A primary challenge in quinary systems is that standard sampling techniques become computationally intractable due to the combinatorial explosion in the configurational space. 
To address this, we restrict the training structures for ternary, quaternary, and quinary compositions to equiatomic stoichiometries, thereby focusing the training data on the central region of the phase diagram most relevant to the Cantor alloy.
This choice is appropriate for resolving SRO near the equiatomic Cantor composition, but it limits direct extrapolation to strongly off-stoichiometric decomposition pathways.
Furthermore, we systematically expand the training set using an iterative active learning strategy.
The initial training pool comprises all five unary structures, together with exhaustive enumerations of the equiatomic binary, ternary, and quaternary configuration spaces in the smallest composition-preserving supercells, namely 2-, 3-, and 4-atom cells, respectively.
For the quinary configurations, we begin with a set of equiatomic structures generated by random sampling of supercells containing up to 25 atoms.
Once the preliminary cluster expansion model is established, we generate additional quinary training structures by performing canonical Monte Carlo (MC) simulations on the existing equiatomic supercells at 1000, 600, and 200~K.
These temperatures are chosen to sample configurations with progressively stronger short-range order, which is expected to develop even within the relatively small supercells used for training.
The energies of the newly sampled configurations are subsequently calculated with DFT and added to the master pool, thereby enriching the next generation of the cluster expansion model with locally ordered environments most relevant to the equiatomic Cantor composition.
The iterative loop is continued until the cross-validation root-mean-square error (RMSE) no longer decreases appreciably.
In total about 3200 training data are generated in the master training pool.

Our cluster expansion considers pair clusters up to the sixth nearest-neighbor (6NN) shell and triplet clusters within the first two NN shells.
The ECIs are determined using a genetic algorithm feature selection by minimizing the RMSE within a 10-fold cross validation scheme, thereby reducing the risk of overfitting and providing a basis for model selection.

To quantify the uncertainty in predicted thermodynamic properties, we train an ensemble of twelve cluster expansion models (\texttt{clex-00} to \texttt{clex-11}) by a random subsampling technique.
Each model is trained on a distinct subset of the master pool of configurations, comprising a fixed set of all unary and binary configurations plus a random 80\% sample of ternary, quaternary, and quinary configurations.
An analysis of the accuracy obtained with the ensemble of twelve models can be found in the Supplementary Information (SI). 
The ensemble yields an average RMSE of 5~meV/atom for the equiatomic quinary systems.

Canonical MC simulations are carried out with a $15\times15\times15$ FCC supercell comprising 13500 atoms. 
The system is initialized at a MC temperature of 1800~K and is then cooled to 10~K with a step of $-10$~K. 
The MC simulations are equilibrated by 10000 sweeps (steps per atom) followed by 20000 sweeps for the production run.
Observables are sampled every 10 MC sweeps to mitigate correlation.
Finite-size effect and statistical uncertainties of the MC simulations are given in the SI. 
Cluster expansions and canonical MC simulations are performed with the \textsc{casm} code~\cite{VanderVen2018}. 

Spin-polarized DFT calculations are performed using the \textsc{vasp} package~\cite{Kresse1996,Kresse1996a,Kresse1999} with the semilocal Perdew-Burke-Ernzerhof (PBE) functional~\cite{Perdew1996}. A kinetic plane-wave cutoff energy of 460~eV is chosen for the projector-augmented wave potentials. The $\mathbf{k}$-space is sampled by a grid density of 1500 per reciprocal atom. 
The initial magnetic moment is 2.0 (Cr, Mn, Ni) and 3.0 (Co, Fe). 
The atomic positions and the unit cell volume are allowed to relax until the forces are below 0.02~eV/\AA.

\section{Results}

\subsection{Pair interaction energies}
Having established the CE Hamiltonian and its uncertainty through the ensemble of models, we first examine the interaction energy of the atomic pairs $\Omega$, also known as the effective pair interaction, through the site basis functions and the ECIs
\begin{equation}\label{eq:omegas}
\Omega(\sigma_i,\sigma_j) = \sum_{\mu,\nu}J_{\mu\nu}(\mathbf{R}) \cdot \Theta_\mu(\sigma_i) \cdot \Theta_{\nu}(\sigma_j),
\end{equation}
where $\sigma_i$ denotes the chemical species occupying site $i$, 
$\Theta_\mu(\sigma)$ is the site basis function $\mu$ for species $\sigma$,
and the ECI $J_{\mu\nu}(\mathbf{R})$ refers to the coupling of basis function $\mu$ on one site and basis function $\nu$ on the other site separated by the distance vector $\mathbf{R}$ defined by the NN shell.
The interaction energy indicates the likelihood of finding an atomic pair in a specific shell, and provides the simplest microscopic view of the chemical driving forces of the SROs.

\begin{figure*}
\includegraphics[width=0.85\textwidth]{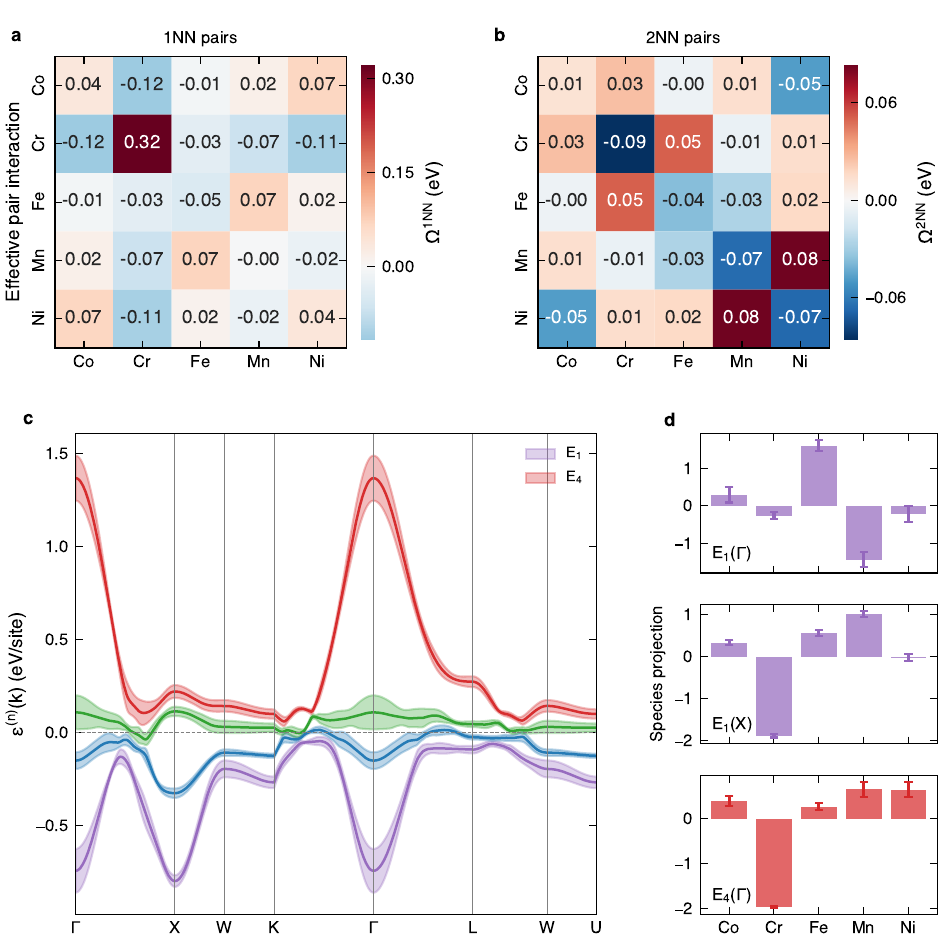}
\caption{\label{fig:omegas_vk}Heatmap of pair interaction energies and concentration wave analysis of the Cantor alloy. (a) and (b) Interaction energies for the atomic pairs in the first two NN shells. (c) Eigenvalues of the pair ECI matrix along the high-symmetry lines in the Brillouin zone. Shaded regions indicate the standard deviation calculated across the ensemble of cluster expansion models. (d) Projected eigenvectors (in the species space) illustrating the atomic site preferences for the three dominating modes. Error bars represent the standard deviation of the projection amplitudes.}
\end{figure*}

The effective pair interactions for the first two NN shells are presented in Fig.~\ref{fig:omegas_vk}(a) and (b).
The complete spatial dependence of the interactions extending up to the 6NN shell is available in Fig.~S2 of the SI.
An outstanding feature of the interaction energy heatmap is the markedly repulsive interaction of the Cr-Cr pair in the 1NN shell [cf.\ Fig.~\ref{fig:omegas_vk}(a)], in contrast to the favorable formation of the heteroatomic Cr pairs such as Cr-Ni and Cr-Co. 
The other atomic pairs show considerably weaker interactions in the 1NN shell.
Meanwhile, pair interactions in the 2NN generally exhibit a sign reversal compared to those in the 1NN shell [cf.\ Fig~\ref{fig:omegas_vk}(b)], a characteristic signature of ordered phases on the FCC lattice.

We also consider the mixing energy for the heteroatomic pairs, which can be expressed in terms of effective pair interactions as
\begin{equation}\label{eq:mix}
\Delta H_\text{mix}(\sigma_i,\sigma_j) = \Omega(\sigma_i,\sigma_j) - \frac{1}{2}\left[ \Omega(\sigma_i,\sigma_i) + \Omega(\sigma_j,\sigma_j) \right].
\end{equation}

The mixing energies of all atomic pairs within the first two nearest neighbors are shown in Fig.~S3.
Predominant mixings are found for the Cr-$X$ pairs in the 1NN shell, whereas the pairs among the other four elements show little to vanishing mixing. 
The mixing energies in the 2NN shell are largely positive, suggesting that heteroatomic pairs are less likely to be found in the 2NN shell compared to homoatomic ones.
Thus far, the pair-interaction analysis suggests a clear ordering tendency in the Cantor alloy at 0~K: the 1NNs are predominantly unlike atoms, whereas the 2NNs consist mainly of like atoms.
This specific alternation of site occupancy is compatible with ordered FCC motifs such as $L1_2$ and $L1_0$, although the full finite-temperature behavior must be assessed with the complete CE Hamiltonian.

\subsection{\label{sec:eci_k}Eigenmode analysis of pair interactions in reciprocal space}
While the real-space ECIs provide a direct measure of local chemical bonding, deducing the global ground state solely from these local parameters is non-trivial, particularly in systems with competing interactions or geometric frustration.
Following the concentration wave theory originally developed by Khachaturyan~\cite{Khatcharuryan1983,Singh2015,Woodgate2022}, we carry out herein an eigenmode analysis of the pair ECIs in reciprocal space, which allows us to survey simultaneously all the periodic arrangements spanning the crystal by calculating the Fourier transform of the real-space ECIs 
\begin{equation}
V_{\mu\nu}(\mathbf{k}) = \sum_\mathbf{R} J_{\mu\nu}(\mathbf{R})e^{-i\mathbf{k}\cdot \mathbf{R}}.
\end{equation}
For a quinary system described by four independent basis functions, $V_{\mu\nu}(\mathbf{k})$ is a $4\times4$ Hermitian matrix.
{\rev{At any $\mathbf{k}$ point in the Brillouin zone, the eigenvalues $\varepsilon^{(n)}(\mathbf{k})$ 
quantify the pair-interaction energy cost or gain associated with a concentration wave of the disordered equiatomic reference, whereas the eigenvectors $v^{(n)}(\mathbf{k})$ define the chemical character of the mode.
Within this convention, negative eigenvalues identify pairwise ordering instabilities of a reference state, whereas positive eigenvalues correspond to concentration waves penalized by the pair interactions.
We note that the negative branches alone are not sufficient to determine the fully ordered ground-state structure, which requires an explicit search over discrete atomic configurations and may involve superpositions of symmetry-related concentration waves.}}

The eigenvalue spectrum in Fig.\ref{fig:omegas_vk}(c) identifies two dominant concentration-wave modes, labeled $E_1$ and $E_4$. 
The dispersion of the lowest branch $E_1(\mathbf{k})$ shows a near-degeneracy between the X point and the $\Gamma$ point. 
A minimum at X corresponds to an instability at a finite wavevector along $\langle 100\rangle$, consistent with $L1_0$/$L1_2$–like ordering tendencies.
Projecting the eigenvector at $E_1$(X) into the species space spanned by the orthonormal Chebyshev polynomials indicates a dominant anti-phase relationship in Cr-$X$ [cf.\ Fig.~\ref{fig:omegas_vk}(d)] as the primary chemical driving force for the $\langle100\rangle$-like ordering.
The dominant Cr character of the X-point instability is qualitatively consistent with experimental observations in related Cr-containing concentrated alloys, where Cr-centered local order that suppresses Cr-Cr nearest neighbors has been linked to $L1_2$ motifs~\cite{Schoenfeld2019,Hsiao2022}.

By contrast, the minimum at $\Gamma$ signals a clustering instability in the long-wavelength limit of composition fluctuations, which, according to the eigenvector analysis, stems primarily from Fe and Mn atoms.
We further point out that the projected eigenvector of the highest branch $E_4(\mathbf{k})$ at $\Gamma$ indicates a strong energetic penalty against Cr clustering.
Altogether, the dispersion spectrum leads to an ordering landscape that is dominated by the competition between the Cr-associated ordering and the clustering of Fe and Mn atoms. 

{\rev{We emphasize that this spectral analysis serves a linearized diagnostic based strictly on the pair ECIs.
It maps the leading pairwise ordering tendencies but does not necessarily delineate a true ground state.}}
To assess the finite-temperature behavior and the role of many-body interactions, we proceed with MC simulations using the full set of ECIs (including triplets).

\subsection{Order-disorder transition}

\begin{figure*}
\includegraphics[width=0.85\textwidth]{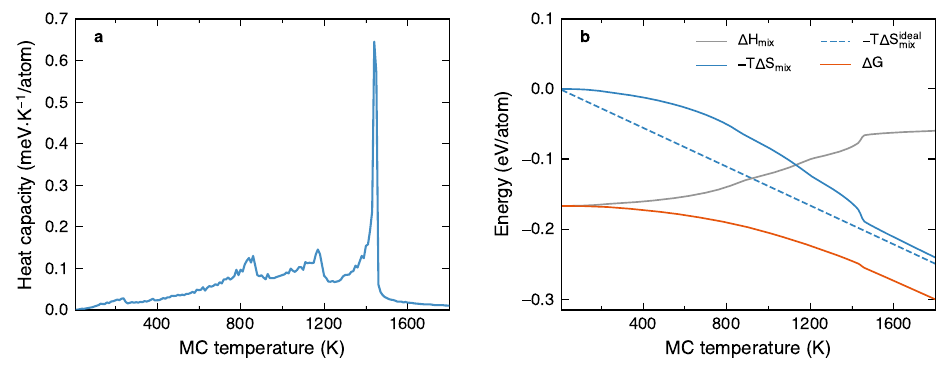}
\caption{\label{fig:mc}Thermodynamic properties of the Cantor alloy with varying temperature (a) Heat capacity vs MC temperature obtained with the representative CE model \texttt{clex-04}. The full ensemble of heat-capacity curves is shown in the Supplementary Information. (b) Enthalpy of mixing, configurational entropy, and Gibbs free energy obtained with the same model.}
\end{figure*}

We now discuss the thermodynamic properties of the Cantor alloy at finite temperatures by performing canonical MC simulations using the ECIs from the CE.
The heat capacity at constant volume is calculated in terms of the variance of the potential energy 
\begin{equation}
C_v(T) = \frac{\langle E^2 \rangle - \langle E \rangle ^2}{k_BT^2}.
\end{equation}
As shown in Fig.~\ref{fig:mc}(a) for the representative model (\texttt{clex-04}), the heat-capacity curve exhibits a primary peak at about $1400$~K, indicative of an ordering transition within the CE Hamiltonian.
From the full set of 12 CE models (see Fig.~S5), the corresponding phase transition temperature
$T_{C_1}$ is found to be $1428 \pm 42$~K (see Table~S1).
\rev{This temperature should not be interpreted as a quantitative equilibrium transition temperature of the real alloy, 
because our MC simulations do not explicitly include vibrational entropy, magnetic interactions, or finite-temperature lattice distortions.}
\rev{Two broader lower-temperature features are observed at $T_{C_2}=1174 \pm 44$~K and $T_{C_3}=711 \pm 146$~K.
The higher degree of uncertainty of $T_{C_3}$ shows the more subtle, model-dependent nature of this low-temperature instability.
These features should not be considered as separate macroscopic phase transitions. 
Instead, they describe changes in local chemical ordering as will be revealed by the SRO eigen-decomposition in Sec.~\ref{sec:sro}.}

Experimental characterization of phase transition is difficult for HEAs, and a precise determination of the phase transition temperature remains largely elusive. Nevertheless, it has been shown that single-phase solid solution can be achieved after annealing at $T_a=1100^{\circ}$C (1373~K)~\cite{LaurentBrocq2015} and 1273~K~\cite{Otto2013}, and various secondary phases have been observed in the range $450 < T_a < 900^{\circ}$C (723--1173~K)~\cite{Otto2016,Schuh2015}. 
While these annealing observations are not direct equilibrium transition temperatures, they are qualitatively consistent with the hierarchy calculated here: a primary high-temperature ordering transition near $T_{C_1}=1428\pm42$~K followed by progressively stronger lower-temperature ordering and clustering tendencies. 

Figure~\ref{fig:mc}(b) shows the thermodynamic properties calculated with one specific CE model (\texttt{clex-04}, see Fig.~S1).
Assuming vanishing entropy for the ordered reference state at very low temperatures near 0~K,
the configurational entropy relative to this reference can be obtained by thermodynamic integration
\begin{equation}
\Delta S_\text{mix}(T) = \int_{0}^{T} \frac{C_v(T')}{T'} dT'.
\end{equation}
\rev{In the finite canonical MC simulations, the heat capacity associated with the primary transition appears as a broadened peak, 
and the thermodynamic integral is numerically well defined.}
Characteristic of solid solutions, $\Delta S_\text{mix}$ approaches the ideal limit of a quinary system ($k_B\ln5$ eV/K$\cdot$atom) at high temperatures.
\rev{Above $T_{C_1}$, this quantity represents the configurational entropy of the high-temperature equiatomic solid-solution state within the fixed-lattice CE Hamiltonian, without vibrational, magnetic, or off-stoichiometric multiphase contributions.}
Concurrently, $\Delta H_\text{mix}$ exhibits an abrupt increase at the primary transition temperature $T_{C_1}$ before reaching a plateau.
The high configurational entropy in this regime is sufficient to offset the mixing enthalpy penalty, thereby promoting the stability of the single-phase quinary alloy.
Overall, the Gibbs free energy of mixing decreases monotonically with temperature, as the free-energy stabilization becomes increasingly entropy dominated at higher temperatures.
As the temperature decreases, the configurational entropy can exhibit large deviations from the ideal value. 
As such, the enthalpy exerts a greater influence on the equilibrium state, and the system minimizes its total Gibbs free energy by lowering its enthalpy via the formation of SRO at the expense of a reduced configurational entropy.

\subsection{\label{sec:sro}Short-range ordering}

\begin{figure*}
\centering
\includegraphics[width=0.85\textwidth]{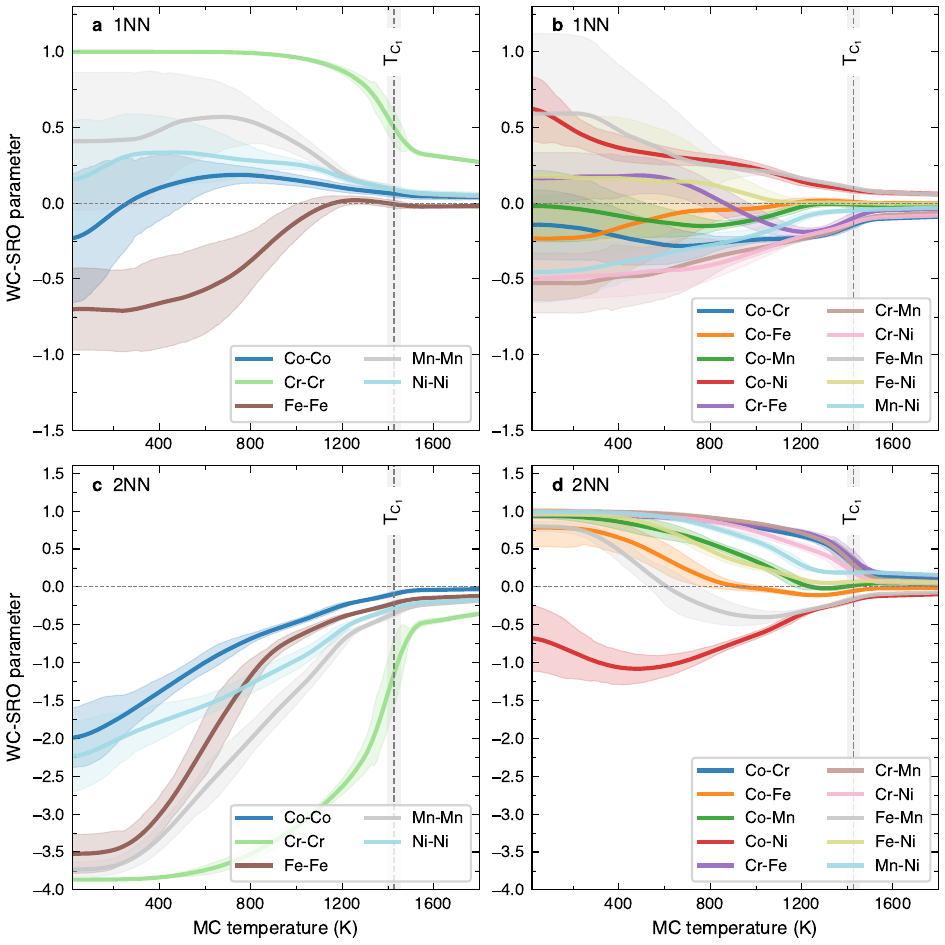}
\caption{\label{fig:sro}Warren-Cowley SRO parameters. (a) and (b) SRO parameters of atomic pairs in the first NN. (c) and (d) SRO parameters in the second NN. The uncertainty of the SROs is indicated by the standard deviation among the results obtained with the model ensembles.}
\end{figure*}

The local ordering of a multicomponent system in the $n$-th NN shell can be quantified by the Warren-Cowley SRO (WC-SRO) parameter
\begin{equation}\label{eq:wc}
\alpha_{ij}^n = 1 - \frac{P^n_{j|i}}{c_j},
\end{equation}
where $P^n_{j|i}$ denotes the conditional probability of finding an atom $j$ in the $n$-th shell given an atom $i$ in the center, and $c_j$ is the global concentration of atom $j$.
The pairwise parameters, $\alpha_{ij}$, are obtained efficiently using a spectral method.
Instead of direct summation in the real space which becomes computationally prohibitive for large supercells, 
we map the atomic site occupancies onto a discrete grid upon which pair correlations are calculated via the fast Fourier transform by exploiting the convolution theorem (see SI for details).
For quinary systems, the WC-SRO parameter ranges from 1 for repulsive pairs to $-4$ for attractive pairs.
\rev{While WC-SRO parameters are often discussed with respect to a disordered reference state, Eq.~\eqref{eq:wc} remains well defined for any fixed-composition configuration, as it is constructed from conditional pair probabilities.}
\rev{Below the primary transition temperature, these parameters should be interpreted as shell-resolved pair correlations of an increasingly ordered state, rather than as small deviations from a random solid solution.}

Starting with the five homoatomic pairs shown in Fig.~\ref{fig:sro}(a) and (c), we see persistent alternating Cr-Cr interactions within the first two NN shells, in line with the strong effective pair interaction energies of Cr-Cr presented in Fig.~\ref{fig:omegas_vk}.
The disfavoring formation of Cr-Cr bonds in medium to high entropy alloys has been widely reported~\cite{Tamm2015,Zhang2017,Ding2018,Zhang2020,Woodgate2022}.
By contrast, Fe-Fe is the only atomic pair that is attractive within the first two NN shells at low to intermediate temperatures.
This is consistent with a low-temperature tendency toward Fe-rich local environments, a phenomenon that has been experimentally observed by energy-dispersive x-ray spectroscopy for the Cantor alloy~\cite{Ding2019}.

The SRO tendency of heteroatomic pairs is mostly consistent with the pairwise interaction energy analysis.
For instance, the favorable formation of Cr-$X$ pairs in the 1NN shell is largely established by their negative WC-SRO values.
However, the Cr-Fe bond is suppressed at low temperatures despite its negative mixing energy.
One possible reason underlying the suppression is the highly competing Fe-Fe pair formation.
Experimentally, the presence of heteroatomic Cr-$X$ bonds at the expense of Cr-Cr depletion in the 1NN has been identified for the ternary CrCoNi alloy by extended x-ray absorption fine structure (EXAFS)~\cite{Zhang2017} and energy-filtered transmission electron microscopy~\cite{Zhang2020}.

While the eigenvalue spectrum at $\Gamma$ identifies a latent instability toward Mn unmixing (cf.\ Fig.~\ref{fig:omegas_vk}),
our MC simulations show no evidence of Mn aggregation.
\rev{This discrepancy is consistent with a state of chemical frustration, in which the driving force for macroscopic Mn unmixing is suppressed by more favorable Cr-Mn local ordering.}
\rev{Conversely, we note that incorporating triplet interactions into the concentration-wave analysis would require a higher-order extension beyond the present formulation. 
Concequently we cannot entirely exclude that from triplet interactions contribute to the observed discrepancy.}

The WC-SRO parameters additionally provide insight into the local pair-correlation patterns present in the Cantor alloy. 
Notably, the 1NN SRO values for multiple Cr-$X$ pairs ($X$=Cr, Ni, Mn) are centered, within statistical uncertainty, around $-\frac{1}{3}$ [cf.\ Fig.~\ref{fig:sro}(b)], which is compatible with $L1_2$/$L1_0$-like correlations.
This structural interpretation is further supported by the positive SRO values close to 1 observed for the 2NN shell [Fig.~\ref{fig:sro}(d)].
Indeed, Sch\"oenfeld et al.\ identified Cr-related L1$_2$-type motifs in the quaternary Cantor derivative CoCrFeNi by diffuse x-ray scattering~\cite{Schoenfeld2019}. 
Although this is not a direct validation for quinary CoCrFeMnNi, it supports the present picture that Cr-centered local ordering provides the dominant microscopic contribution to the primary instability.

Overall, the temperature-dependent WC-SRO parameters indicate that the primary transition near 1400~K is predominantly associated with the order-disorder behavior of the Cr-Cr pairs, whereas the lower-temperature feature near 800~K is consistent with an Fe-rich clustering instability.
However, the SRO parameters alone do not fully resolve the microscopic origin of the secondary ordering instability near 1200~K.

\begin{figure*}
\centering
\includegraphics[width=0.85\textwidth]{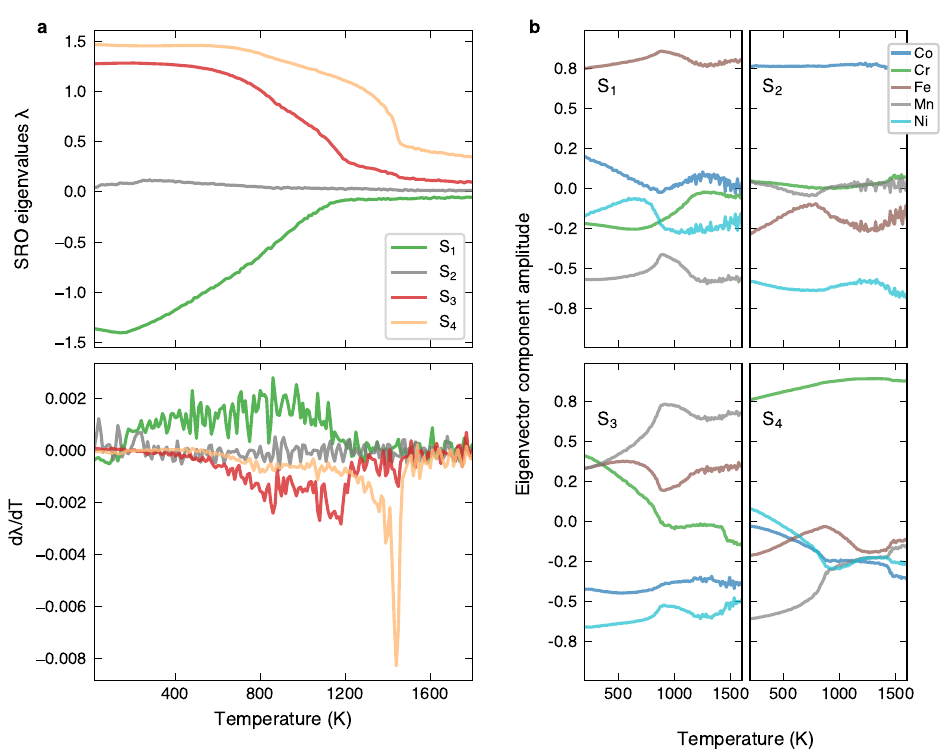}
\caption{\label{fig:eig}Eigen-decomposition of WC-SRO parameters in the 1NN shell. (a) Eigenvalues of the four independent ordering modes ($S_1$ to $S_4$) and their derivatives with respect to the temperature. (b) Eigenvector component amplitude as a function of temperature. The results are obtained with the \texttt{clex-04} model.}
\end{figure*}

To elucidate the microscopic origins of the observed thermal features and their instabilities,
we proceed with an eigen-decomposition analysis of the WC-SRO parameters.  
This approach extends the concentration wave theory in Sec.~\ref{sec:eci_k} to real-space pair correlations.
By diagonalizing the $5\times5$ SRO matrix, we analyze the four non-trivial chemical modes of the equiatomic quinary system; the fifth mode is the concentration-conserving null mode.
Within this framework, the eigenvalues $\lambda^{(n)}$ quantify the magnitude and sign of each fluctuation mode within the $n$-th NN shell, while the eigenvectors $v_i^{(n)}$ represent the orthogonal modes obtained from the eigen-decomposition of the shell-resolved SRO matrix.
Physically, these real-space normal modes are the local projections of the global concentration-wave instabilities identified in reciprocal space.
To distinguish them from the reciprocal-space concentration-wave branches $E_i(\mathbf{k})$, we denote the four real-space SRO modes by $S_1$ to $S_4$.
Within this constrained subspace, the SRO parameters can be represented by the spectral decomposition $\alpha_{ij} = \sum_n \lambda^{(n)} v_i^{(n)} v_j^{(n)}$.
\rev{Note that the SRO eigen-decomposition below $T_{C_1}$ tracks dominant local chemical correlation modes, rather than linear-response fluctuations around a disordered state.}
To identify the temperatures at which these modes evolve most rapidly,
we also calculate the temperature derivative of the eigenvalues, $d\lambda/dT$. 
The extrema in $d\lambda/dT$ provide convenient indicators of the temperature at which the corresponding mode changes most rapidly.
 
Figure~\ref{fig:eig} presents the evolution of the 1NN SRO eigenvalues, their derivatives, and the associated eigenvectors (in the elemental basis) as a function of temperature obtained with the representative \texttt{clex-04} model, which is used here to illustrate the mode-resolved behavior.
An analogous decomposition carried out for all 12 \texttt{clex} models yields the same qualitative ordering hierarchy and closely similar eigenvector character (see Fig.~S8), indicating that the 1NN SRO mode assignment is robust within the CE ensemble.
\rev{The primary ordering at $T_{C_1}\approx1400$~K is most strongly associated with the $S_4$ eigenmode, where the eigenvector is dominated by the Cr component [cf.\ Fig.~\ref{fig:eig}(b)].}
The unfavorable Cr-Cr interaction indicates a strong tendency for Cr to order on a specific sublattice against the solid solution as the system cools down below $T_{C_1}$.
    
The broader lower-temperature feature near $T_{C_2}\approx1200$~K coincides with the onset of the eigenvalue derivative associated with the $S_3$ eigenmode, identifying a secondary ordering instability.
An analysis of the eigenvector components shows that the $S_3$ mode is not merely a binary interaction but a collective sublattice ordering involving all the elements other than Cr.
In particular, the eigenvector signs suggest a structured pattern stemming from the (Co,Ni)-(Fe,Mn) ordering, compatible with the $L1_2$/$L1_0$-like correlations inferred from the earlier concentration-wave and real-space WC-SRO analyses.
\rev{To further assess this interpretation, we calculate diffuse scattering maps and the mode-projected structure factor from the MC snapshots (see Section S7 of the SI). The emergence of enhanced intensity at the FCC X points provides a direct reciprocal-space signature of $L1_2$/$L1_0$-like ordering correlations associated with the $S_3$ chemical contrast.}
The inferred ordering tendency is chemically plausible in light of known ordering tendencies in relevant binaries, notably Ni$_3$Mn ($L1_2$)~\cite{Averbach1951}, FeNi$_3$ ($L1_2$)~\cite{Liu2016}, and FeNi ($L1_0$)~\cite{Bordeaux2016}.

The lower-temperature feature near $T_{C_3}\approx800$~K is mostly associated with the $S_1$ eigenmode, indicating a clustering instability.
The negative sign of its eigenvalue is consistent with a clustering tendency.
The mode is characterized largely by the Fe component, and to some extent, by Mn and Cr albeit with an opposite sign.
\rev{This low-temperature feature thus shows an increased tendency toward Fe-rich local environments at the expense of the depletion of Mn and Cr atoms. It should be emphasized that this clustering mode describes a local ordering tendency within the equiatomic solid
  solution, rather than macroscopic composition partitioning into secondary phases, which is restricted by the canonical simulation setup.}

We note that a parallel eigen-decomposition of the 2NN SRO matrix yields consistent chemical normal modes (cf.\ Fig.~S9). 
In particular, the ordering modes (e.g., $S_3$ and $S_4$) exhibit an inverted eigenvalue sign relative to the 1NN shell. 
This sign reversal is consistent with the expected projection of a $\langle100\rangle$ concentration wave, confirming the extended phase coherence of the alternating atomic layers.
\rev{Together with the reciprocal-space X-point instability and the 1NN/2NN WC-SRO pattern, this supports the assignment of $L1_2$/$L1_0$-like correlations.}

\section{Conclusion}
Our statistical analysis of atomic configurations, utilizing the cluster expansion method augmented by an eigen-decomposition of SRO parameters, suggests a hierarchy of ordering and clustering instabilities in the Cantor alloy.
We find that the primary ordering behavior is driven by repulsive Cr-Cr interactions, which promote robust Cr-centered local motifs compatible with $L1_2$/$L1_0$-like correlations and emerge near the primary ordering temperature.
By analyzing the temperature derivatives of the SRO eigenvalues, we identify distinct consecutive tendencies within this Cr-ordered matrix: a secondary collective ordering mode involving the non-Cr components, followed by a lower-temperature clustering mode consistent with the emergence of Fe-rich local environments.
\rev{It is important to note that, within the fixed equiatomic composition, our predictions do not account for macroscopic phase separation.}

Methodologically, our computational framework, built upon cluster expansion and SRO eigen-decomposition, provides a versatile scheme for analyzing overlapping ordering patterns in multicomponent systems.
\rev{Beyond established Cr-related SRO trends, the present analysis resolves a temperature-dependent hierarchy of collective ordering modes in the equiatomic quinary alloy.}
The predicted ordering hierarchy is qualitatively consistent with experiment: high-temperature annealing retains a single-phase solid solution, lower-temperature annealing promotes secondary phases, and diffuse-scattering measurements in the related CoCrFeNi system reveal Cr-centered local motifs compatible with the present Cr-driven ordering picture~\cite{LaurentBrocq2015,Otto2013,Otto2016,Schuh2015,Schoenfeld2019}. 
In addition, the low-temperature Fe clustering tendency is qualitatively consistent with the local Fe-rich groups reported by atomic-resolution chemical mapping in the Cantor alloy~\cite{Ding2019}.
To further refine the predictive accuracy, future work should extend the training set toward off-equiatomic configurations, address the limitations of the fixed-lattice approximation, account for vibrational contributions, and explicitly model magnetic interactions---a factor recognized in recent literature as critical to the local ordering of high-entropy systems~\cite{Walsh2021,Zhu2024}.

\section*{Acknowlegements}
We acknowledge the computational resources provided by the supercomputing facilities of UCLouvain (CISM/UCL) and Consortium des Equipements de Calcul Intensif en F\'ed\'eration Wallonie-Bruxelles (C\'ECI) funded by the Fond de la Recherche Scientifique de Belgique (F.R.S.-FNRS) under convention 2.5020.11 and by the Walloon Region. G.-M.R.\ is Research Director of the Fonds de la Recherche Scientifique - FNRS. G.H.\ acknowledges support from the U.S. Department of Energy, Office of Science, Office of Basic Energy Sciences Established Program to Stimulate Competitive Research (EPSCoR) program under Award Number DE-SC-0021347.

\section*{Data availability}
The DFT training data upon which the cluster expansion Hamiltonians are built are available at \url{https://doi.org/10.5281/zenodo.18967881}.

\bibliography{main}

\end{document}